\documentclass[10pt,conference]{IEEEtran}
\IEEEoverridecommandlockouts
\usepackage[final]{graphicx}
\usepackage{times}
\usepackage{amsfonts}
\usepackage{amsthm}
\usepackage{epsfig}
\usepackage{amssymb}
\usepackage{amsmath}
\usepackage{citesort}
\usepackage{multirow}
\usepackage{csquotes}
\usepackage{psfrag}
\usepackage{balance}
\usepackage{pgfplots}
\usepackage{algorithm}
\usepackage[noend]{algorithmic}
\usepackage{comment}
\usepackage[binary-units=true]{siunitx}

\newcommand\T{\rule{0pt}{2.6ex}}
\newcommand\B{\rule[-1.2ex]{0pt}{0pt}}
\DeclareMathOperator*{\argmin}{arg\,min}
\DeclareMathOperator{\conv}{conv}

\begin{document}
\title{Efficient Maximum-Likelihood Decoding of Linear Block Codes on Binary Memoryless Channels}

\author{\IEEEauthorblockN{Michael Helmling\IEEEauthorrefmark{2}, Eirik Rosnes\IEEEauthorrefmark{4}, Stefan Ruzika\IEEEauthorrefmark{2}, and Stefan Scholl\IEEEauthorrefmark{3}}\\
\vspace{-4mm}\IEEEauthorblockA{\IEEEauthorrefmark{2}Mathematical Institute, University of Koblenz-Landau,
56070 Koblenz, Germany \\
Email: \{helmling, ruzika\}@uni-koblenz.de}
\IEEEauthorblockA{\IEEEauthorrefmark{4}Department of Informatics,
University of Bergen, N-5020 Bergen, Norway, and the Simula Research Lab.\\ Email:
eirik@ii.uib.no}
\IEEEauthorblockA{\IEEEauthorrefmark{3}Microelectronic Systems Design Research Group,
University of Kaiserslautern, 67653 Kaiserslautern, Germany \\ Email: scholl@eit.uni-kl.de}
\thanks{This work was partially funded by the DFG (grant RU-1524/2-1) and by DAAD / RCN (grant 54565400 within the German-Norwegian Collaborative Research Support Scheme). }}%
\maketitle

\begin{abstract}
In this work, we consider efficient maximum-likelihood decoding of linear block codes for small-to-moderate block lengths. The presented approach is a branch-and-bound algorithm using the cutting-plane approach of Zhang and Siegel (\emph{IEEE Trans.\ Inf.\ Theory}, 2012) for obtaining lower bounds. We have compared our proposed algorithm  to the state-of-the-art commercial integer program  solver CPLEX, and for all considered codes our approach is faster for both low and high signal-to-noise ratios. For instance, for the benchmark $(155,64)$ Tanner code our algorithm is more than 11 times as fast as CPLEX for an SNR of \SI{1.0}{\decibel} on the additive white Gaussian noise channel. By a small modification, our algorithm can be used to calculate the minimum distance, which we have again verified to be much faster than using the CPLEX solver. 
\end{abstract}
\section{Introduction} \label{sec:intro}

Determining the optimal decoding behavior of error-correcting codes is of significant importance, e.\,g., to benchmark different coding schemes. When no \emph{a priori} information on the transmitted codeword is known, maximum-likelihood decoding (MLD) is an optimal decoding strategy. It is known that this problem is NP-hard in general \cite{ber78} such that its complexity grows exponentially in the block length of the code, unless $\text{P}=\text{NP}$. Currently, the best known approach for general block codes is to use a state-of-the-art (commercial) integer program (IP) solver (see \cite{Tanatmis+10NumericalComparison}) like CPLEX \cite{CPLEX126}.

In this work, we present a branch-and-bound approach for efficient MLD of linear block codes. 
The problem of MLD is closely related to that of calculating the minimum distance of a code, which has attracted some attention recently. For instance, in \cite{ros12,ros09}, Rosnes \emph{et al.} proposed an efficient branch-and-bound algorithm to determine all low-weight stopping sets/codewords in a low-density parity-check (LDPC) code. Although the two problems are similar, the bounding step in the algorithm from \cite{ros12,ros09} cannot efficiently be adapted to the scenario of MLD. Conversely, however, the algorithm presented here can also calculate the minimum distance, and our numerical experiments show that this is very efficient compared to CPLEX.  

Linear programming (LP) decoding of binary linear codes, as first introduced by Feldman \emph{et al.} in \cite{fel05}, approximates MLD by relaxing the decoding problem into an easier to solve LP problem. The LP problem contains a set of linear 
inequalities that are derived from the parity-check constraints of a (redundant) parity-check matrix representing the code. As shown in \cite{tag07}, these constraints can iteratively and adaptively be added to the decoding problem, which significantly reduces the overall complexity of LP decoding. Elaborating on this idea, Zhang and Siegel \cite{zha12} proposed an efficient search algorithm for new \emph{violated} (redundant) parity-check constraints (or ``cuts'') that tighten the decoding polytope. 
Depending on the structure of the underlying code, for some codes, this ``cutting-plane'' LP decoding algorithm performs close to MLD for high signal-to-noise ratios (SNRs) on the additive white Gaussian noise (AWGN) channel, although for most codes, e.\,g., the $(155,64)$ Tanner code  \cite{tan01}, there is still a gap in decoding performance to MLD \cite{zha12}. For lower values of the SNR, there could be a significant performance degradation with respect to MLD.

The algorithm proposed in this work closes that gap by using the cutting-plane algorithm for lower bounds and the well-known sum-product (SP) decoder \cite{ksc01} with order-$i$ re-encoding \cite{fos95} for upper bounds within a sophisticated branch-and-bound framework such that the output always and provably is the maximum-likelihood (ML) codeword. Our numerical study in Section~\ref{sec:results} shows that it is much faster than CPLEX for all codes under consideration, and moreover is able to decode some of the codes on which CPLEX fails completely.

\section{Notation and Background} \label{sec:basics}
This section establishes some basic definitions and results needed for the rest of the paper.

Let $\mathcal{C}$ denote a binary linear code of length $n$ represented by an $m \times n$ parity-check matrix $\mathbf{H}$. The code is used on a binary-input memoryless output-symmetric channel with input $\mathbf{c}=(c_0,\dots,c_{n-1}) \in \mathcal{C}$ and channel output denoted by the length-$n$ vector $\mathbf{r}=(r_0,\dots,r_{n-1})$. 
The ML decoder can be described by the following optimization problem \cite{hel12}:
\begin{equation} \label{eq:ML}
\hat{\mathbf{c}}_{\rm ML}=\argmin_{\mathbf{c} \in \mathcal{C}} \psi_{\boldsymbol{\lambda}}(\mathbf{c}) = \argmin_{\mathbf{c} \in \conv(\mathcal{C})} \psi_{\boldsymbol{\lambda}}(\mathbf{c})
\end{equation}
where $\psi_{\boldsymbol{\lambda}}(\mathbf{c}) = \boldsymbol{\lambda} \cdot \mathbf{c}^T$ and $(\cdot)^T$ denotes the transpose of its argument, $\boldsymbol{\lambda}=(\lambda_0,\dots,\lambda_{n-1})$ is a vector of log-likelihood ratios (LLRs) defined by
\begin{equation} \notag
\lambda_i = \log \left( \frac{{\rm {Pr}}(r_i|c_i=0)}{{\rm {Pr}}(r_i|c_i=1)} \right)
\end{equation}
for all $i$, $0 \leq i \leq n-1$, 
and $\conv(\mathcal{C})$ is the \emph{convex hull} of $\mathcal{C}$ in $\mathbb{R}^n$, where $\mathbb{R}$ denotes the real numbers. 
The MLD problem in (\ref{eq:ML}) can be formulated as an IP which in general is an NP-hard problem. As an approximation to MLD, Feldman \emph{et al.} \cite{fel05} relaxed the codeword polytope $\conv(\mathcal{C})$ in the following way.

Define
\begin{equation} \notag
\mathcal{C}_j = \{ \mathbf{c} \in \{0,1\}^n\,|\, \mathbf{h}_j \cdot \mathbf{c}^T = 0 \}
\end{equation}
where $\mathbf{h}_j = (h_{j,0},\dots,h_{j,n-1})$ is the $j$th row of the parity-check matrix $\mathbf{H}$ and $0 \leq j \leq m-1$. Furthermore, let $\conv(\mathcal{C}_j)$ denote the convex hull of $\mathcal{C}_j$ in $\mathbb{R}^n$. The \emph{fundamental polytope} $\mathcal{P}(\mathbf{H})$ of the parity-check matrix $\mathbf{H}$ is defined as \cite{koe05}
\begin{equation} \label{eq:fundapoly}
\mathcal{P}(\mathbf{H}) = \bigcap\nolimits_{j=0}^{m-1} \conv(\mathcal{C}_j).
\end{equation}
The MLD problem in (\ref{eq:ML}) can now be relaxed to
\begin{equation} \notag
\hat{\mathbf{p}}_{\rm LP}= \argmin_{\mathbf{p} \in \mathcal{P}(\mathbf{H})} \psi_{\boldsymbol{\lambda}}(\mathbf{p})
\end{equation}
where the solution, by definition, is a \emph{pseudocodeword} with  fractional entries in general. Note that the LP decoder has the \emph{ML certificate} property, which means that in case $\hat{\mathbf{p}}_{\rm LP}$ is integral, it is an optimal solution to \eqref{eq:ML}.

For each row $\mathbf{h}_j$, $0 \leq j \leq m-1$, in the matrix $\mathbf{H}$ the linear inequalities behind the fundamental polytope in (\ref{eq:fundapoly}) are 
\begin{equation} \label{eq:parityineq}
\sum_{i \in \mathcal{V}} p_i  - \!\!\sum_{i \in \mathcal{N}(j) \setminus \mathcal{V}} \!\!p_i \leq |\mathcal{V}|-1,\ \text{for all odd-sized }\mathcal{V} \subseteq \mathcal{N}(j)
\end{equation}
where $\mathcal{N}(j) = \{i : h_{j,i}=1,\ 0 \leq i \leq n-1 \}$.

For a given row $\mathbf{h}_j$ of a parity-check matrix $\mathbf{H}$ and a vector $\mathbf{p} \in [0,1]^n$: If there exists an odd set $\mathcal{V} \subseteq \mathcal{N}(j)$ such that the corresponding inequality from (\ref{eq:parityineq}) does not hold, then we say that the $j$th parity-check constraint induces a \emph{cut} at $\mathbf{p}$.

Central to our branch-and-bound algorithm is the concept of \emph{constraint sets}. A constraint set $F$ is a set $ \{(\rho_i,c_{\rho_i}) :
 c_{\rho_i} \in \{0,1\}\;\forall \rho_i \in \Gamma \}$, where $\Gamma\subseteq\{0,\ldots,n-1\}$. If $(\rho_i,c_{\rho_i})$ is a constraint, then position $\rho_i$ is said to be $c_{\rho_i}$-constrained, which means that position $\rho_i$ is committed to the value $c_{\rho_i}$ in a codeword, while positions not in $F$ are uncommitted. 

Let $\mathcal{C}^{(F)}$ denote the subset of codewords from $\mathcal{C}$ consistent with the constraint set $F$. Then, we can define
\begin{equation} \notag
\psi^{(F)}_{\rm min, \boldsymbol{\lambda}} = \min_{\mathbf{c} \in \mathcal{C}^{(F)}} \psi_{\boldsymbol{\lambda}}(\mathbf{c})
\end{equation}
as the minimum value of the objective function for codewords consistent with $F$. In the following, $\bar{\psi}^{(F)}_{\rm min, \boldsymbol{\lambda}}$ will denote any lower bound on $\psi^{(F)}_{\rm min, \boldsymbol{\lambda}}$. Also, a constraint set $F$ is said to be \emph{valid} if  $\mathcal{C}^{(F)}$ is nonempty.

Our proposed branch-and-bound MLD algorithm relies heavily on tight lower bounds on $\psi^{(F)}_{\rm min, \boldsymbol{\lambda}}$, which are provided by an LP-based decoding algorithm by Zhang and Siegel \cite{zha12}. We briefly describe this algorithm, denoted as the ZS decoding algorithm, below in Section~\ref{sec:ZS}.

\subsection{Zhang and Siegel's LP-Based Decoding Algorithm} \label{sec:ZS}
The ZS decoding algorithm is based on adaptive LP decoding as described in \cite{tag07} and incorporates an efficient cut-search algorithm, as described in \cite{zha12}. First the LP problem is initialized with the box constraints. $\dagger$Solve the LP problem to get an optimal solution $\mathbf{p}^*$. If $\mathbf{p}^*$ is integral, then terminate the algorithm and return the ML codeword $\mathbf{p}^*$. Otherwise, the cut-search algorithm (Algorithm~1 in \cite{zha12}) is applied to each row of the parity-check matrix of the code.  If at least one is found, add all found cuts into the LP problem and repeat the procedure from $\dagger$. Otherwise, search for cuts from redundant parity-checks. To this end, reduce $\mathbf{H}$ by Gaussian elimination to \emph{reduced row echelon} form, where the columns of $\mathbf{H}$ are processed in the order of the \enquote{fractionality} (i.\,e., closeness to $\frac12$) of the corresponding coordinate of $\mathbf{p}^*$.
Now, the cut-search algorithm is applied to each row of the obtained modified matrix $\tilde{\mathbf{H}}$. If no cut is found, then terminate. Otherwise, add all found cuts 
into the LP problem as constraints and repeat the procedure from $\dagger$. The algorithm above has been detailed in Algorithm~2 in \cite{zha12}, and we refer the interested reader to  \cite{zha12} for further details.

\section{Basic Branch-and-Bound Algorithm} \label{sec:basicalg}


%
%
Our proposed algorithm is a branch-and-bound algorithm on constraint sets and uses the ZS decoding algorithm, as briefly described above, as a basic component in the bounding step. Thus, there is a one-to-one correspondence between the nodes in the search tree and constraint sets. 
In the following, when we speak about the \emph{left} and \emph{right} child constraint set, denoted by $F^{\downarrow 0}$ and $F^{\downarrow 1}$, respectively, we mean the constraint set of the left and right child nodes in the search tree.

Now, to each constraint set $F$, we associate three real numbers $\bar{\psi}_{\rm min, \boldsymbol{\lambda}}^{(F)}$,  $\bar{\psi}_{\rm min, \boldsymbol{\lambda}}^{(F){\downarrow 0}}$, and  $\bar{\psi}_{\rm min, \boldsymbol{\lambda}}^{(F){\downarrow 1}}$  which are \emph{current} lower bounds on $\psi_{\rm min,\boldsymbol{\lambda}}^{(F)}$,  $\psi_{\rm min,\boldsymbol{\lambda}}^{(F^{\downarrow 0})}$, and  $\psi_{\rm min,\boldsymbol{\lambda}}^{(F^{\downarrow 1})}$, respectively. When a constraint set is created, these values are initiated to $-\infty$.

The algorithm maintains a list $L$ of active constraint sets which is initiated with the unconstrained set $\emptyset$. In each iteration, a constraint set $F$ is selected from the list according to the node selection rule (see below). A valid codeword, i.\,e., a feasible solution of the IP, is generated by decoding the LLR vector (where the constraints imposed by $F$ are enforced by altering the according LLR values to $\pm \infty$) using the SP algorithm \cite{ksc01} 
with order-$i$ re-encoding \cite{fos95}, for some integer $i$,  as a post-processing step. The upper bound on the objective function decreases if the decoder output has a lower objective function value than the previous best candidate.
Afterwards a lower bound on $\psi^{(F)}_{{\rm min},\boldsymbol{\lambda}}$ is computed by running the ZS algorithm, where the variables contained in $F$ are fixed to their corresponding values. If an integral solution is returned, i.\,e., a pseudocodeword  with no fractional coordinates, it is considered another candidate codeword in the same way as above. Otherwise, and if the computed lower bound 
is less than the  current upper bound $\tau$, the algorithm branches on a fractional position, selected by the branching rule (see below), of the pseudocodeword by adding two constraint sets, namely $F$ augmented by the chosen branching position fixed to $0$ and $1$, respectively, to the list of active nodes.
Then, the next set is chosen from $L$ until one of the termination criteria in Step~\ref{line:mainloop} of Algorithm~\ref{alg:sse} (which gives a formal description of the overall algorithm) is fulfilled.  Note that the computations to produce  a lower bound on $\psi^{(F)}_{{\rm min},\boldsymbol{\lambda}}$ 
for a given constraint set $F$ are collected into Algorithm~\ref{alg:ZS}, denoted by LUBD. 

\subsection{Bounding Step}

The complexity of Algorithm~\ref{alg:sse} depends heavily on the tightness of the lower bounds computed in Step~\ref{line:lubd} (from Algorithm~\ref{alg:ZS}), i.\,e., on how close $\psi_{\boldsymbol{\lambda}}(\hat{\mathbf{p}})$ is to the value $\psi^{(F)}_{\rm min, \boldsymbol{\lambda}}$. To find the best pseudocodeword $\hat{\mathbf{p}}$, we have used the procedure detailed in Algorithm~\ref{alg:ZS}.

Note that $\min\{\psi^{(F \downarrow0)}_{\min,\boldsymbol{\lambda}},\psi^{(F\downarrow1)}_{\min,\boldsymbol{\lambda}}\} = \psi^{(F)}_{\min,\boldsymbol{\lambda}}$ for any node (constraint set) $F$. This allows us to update the current lower bound $\bar{\psi}^{(F)}_{\min,\boldsymbol{\lambda}}$ of $F$ (and, recursively, also the 
ancestors of $F$), potentially \emph{increasing} its value, once both of its children have been processed (see Steps \ref{line:updbd}--\ref{line:updbd-end} of Algorithm~\ref{alg:sse}). Tightening the bounds in the search tree is important for decreasing the complexity of the algorithm because nodes whose lower bound exceeds the objective value of the currently best candidate solution (i.\,e., the current upper bound) can be skipped, thereby reducing the search space.

\subsection{Branching Step}

We have used the following simple branching rule to select the position $p$ in Step~\ref{line:branch} of Algorithm~\ref{alg:sse}: Take an unconstrained position where the corresponding entry in the decoded pseudocodeword $\hat{\mathbf{p}}$ is closest to $\frac12$. This simple procedure seems to work very well in practice.

\subsection{The Processing Order of the List $L$}\label{sec:nodeselection}
The node selection rule, i.\,e., the method by which a constraint set $F$ is selected from $L$ in Step~\ref{line:selectnode} of Algorithm~\ref{alg:sse}, has great influence on the overall complexity. The most common schemes are \emph{depth-first search}, according to LIFO (last in -- first out) processing of $L$, and \emph{breadth-first search}, where $L$ is processed in FIFO (first in -- first out) fashion. Another popular method, called \emph{best-bound search}, selects the next constraint set by $F' = \argmin_{F \in L} \bar{\psi}_{\min,\boldsymbol{\lambda}}^{(F)}$, with the goal of tightening the overall lower bound as fast as possible.

In our experiments, the following mixed strategy has proven to be most efficient. Apply depth-first processing in general, but every $M$ iterations, for a fixed integer $M$, and only if $\bar{\psi}_{\min,\boldsymbol{\lambda}}^{(F)} < \tau-\delta$ for a fixed $\delta>0$, where $F$ is the constraint set from the previous iteration, select the next node by the best-bound rule above.

\begin{algorithm}[t] 
\caption{Maximum-Likelihood Decoding (MLD)}
\label{alg:sse}
{\bf Input:} The received LLR vector $\boldsymbol{\lambda}$ and the order $i$ of re-encoding.\\
{\bf Output:} An ML decoded codeword $\hat{\mathbf{c}}_{\rm ML}$.\\[-2.2ex]
\begin{algorithmic}[1]
\STATE Initialize $\tau \leftarrow \infty$ and $L \leftarrow \{ \emptyset \}$
\WHILE {$L \neq \emptyset$ and $\bar{\psi}^{(\emptyset)}_{\min,\boldsymbol{\lambda}} < \tau$} \label{line:mainloop}
  \STATE Choose and remove a constraint set $F$ from $L$.\label{line:selectnode}
  \IF{$F$ is valid and $\bar\psi^{(F)}_{\min,\boldsymbol{\lambda}} < \tau$}\label{line:prune1}
    \STATE let $(\hat{\mathbf{c}},\hat{\mathbf{p}}) \leftarrow \text{LUBD}(\boldsymbol{\lambda},F)$\label{line:lubd}
    \IF{$\psi_{\boldsymbol{\lambda}}(\hat{\mathbf{c}}) < \tau$}
      \STATE let $\hat{\mathbf{c}}_{\rm ML} \leftarrow \hat{\mathbf{c}}$ and $\tau \leftarrow \psi_{\boldsymbol{\lambda}}(\hat{\mathbf{c}})$
    \ENDIF
    \STATE $\bar{\psi}^{(F)}_{\min,\boldsymbol{\lambda}} \leftarrow \psi_{\boldsymbol{\lambda}}(\hat{\mathbf{p}})$
    \IF{$\hat{\mathbf{p}}$ is integral}\label{line:integrality}
      \IF{$\psi_{\boldsymbol{\lambda}}(\hat{\mathbf{p}}) < \tau$}
        \STATE let $\hat{\mathbf{c}}_{\rm ML} \leftarrow \hat{\mathbf{p}}$ and $\tau \leftarrow \psi_{\boldsymbol{\lambda}}(\hat{\mathbf{p}})$
      \ENDIF
    \ELSIF{$\psi_{\boldsymbol{\lambda}}(\hat{\mathbf{p}}) < \tau$}\label{line:prune2}
      \STATE\label{line:branch} choose an unconstrained position $p$ based on $\hat{\mathbf{p}}$, construct two new constraint sets $F'=F\cup\{(p,0)\}$ and $F''=F\cup\{(p,1)\}$, and append them to $L$.
    \ENDIF
  \ENDIF
  \IF{$F \neq \emptyset$}\label{line:updbd}
    \STATE determine (the unique) $\tilde{F}$ such that $F=\tilde F^{\downarrow i}$, where $i=0$ or $1$, and update parent bounds as follows:\\
    \STATE $\bar{\psi}^{(\tilde F)\downarrow i}_{\min,\boldsymbol{\lambda}} \leftarrow \max\{\bar{\psi}^{(\tilde F)\downarrow i}_{\min,\boldsymbol{\lambda}}, \bar{\psi}^{(F)}_{\min,\boldsymbol{\lambda}}\}$\\
   \STATE $\bar{\psi}^{(\tilde F)}_{\min,\boldsymbol{\lambda}} \leftarrow \max\{\bar{\psi}^{(\tilde F)}_{\min,\boldsymbol{\lambda}},\min \{\bar{\psi}^{(\tilde F)\downarrow 0}_{\min,\boldsymbol{\lambda}},\bar{\psi}^{(\tilde F)\downarrow1}_{\min,\boldsymbol{\lambda}}\}\}$\\
   \STATE If $\bar{\psi}^{(\tilde F)}_{\min,\boldsymbol{\lambda}}$ increased in the previous step, recurse to Step~\ref{line:updbd} with $F$ replaced by $\tilde F$\label{line:updbd-end}.
 \ENDIF
\ENDWHILE
\STATE Return $\hat{\mathbf{c}}_{\rm ML}$.
\end{algorithmic}
\end{algorithm} 

\begin{algorithm}[t] 
\caption{Lower and Upper Bound Algorithm (LUBD)}
\label{alg:ZS}
{\bf Input:} The received LLR vector $\boldsymbol{\lambda}$ and a constraint set $F$.\\
{\bf Output:} The pair $(\hat{\mathbf{c}},\hat{\mathbf{p}})$.\\[-2.2ex]
\begin{algorithmic}[1]
\STATE Perform SP decoding with order-$i$ re-encoding on an LLR vector constrained according to $F$ as follows:
\begin{itemize}
\item $+\infty$ for positions corresponding to $0$-constraints.
\item $-\infty$ for positions corresponding to $1$-constraints.
\item The original channel LLRs for  positions not in $F$.
\end{itemize}
The resulting decoded codeword is denoted by $\hat{\mathbf{c}}$.
\STATE Perform ZS decoding on the received LLR vector $\boldsymbol{\lambda}$ with equality constraints according to $F$. Denote the decoded pseudocodeword by $\hat{\mathbf{p}}$.
\STATE Return the pair $(\hat{\mathbf{c}},\hat{\mathbf{p}})$.
\end{algorithmic}
\end{algorithm}

\section{Improvements} \label{sec:improvements}
In this section, we present some improvements to the basic algorithm from Section~\ref{sec:basicalg}. 

\subsection{Tuning the ZS Algorithm for Adaptive LP Decoding} \label{sec:constraints}

A linear inequality constraint of the general form $\mathbf{a}\cdot \mathbf{x}^T  \leq b$, where $\mathbf{a}$ and $b$ are constants, is called \emph{active} at the point $\mathbf{x}^*$ if it holds with equality for $\mathbf{x} = \mathbf{x}^*$. Otherwise, it is called \emph{inactive}. For an LP problem with a set of linear inequality constraints, the optimal solution $\mathbf{x}_{\text{LP}}$ is a \emph{vertex} of the polytope formed by the hyperplanes of all constraints that are active at $\mathbf{x}_{\text{LP}}$. Thus, constraints inactive at $\mathbf{x}_{\text{LP}}$ can be removed without changing the optimal solution.

The ZS decoding algorithm uses adaptive LP decoding, which implies that a lot of linear programs (of increasing size in the number of constraints) are solved successively. Consequently, a simple way to reduce the overall complexity is to remove inactive constraints from time to time.

Our implementation uses the \emph{dual simplex method} for solving LP problems, which is very effective in the case of iteratively added constraints by employing a warm-start technique that reuses the basis information of the previously optimal solution (see, e.\,g., \cite{faigle10} for details). The removal of constraints, however, is expensive because afterwards a new simplex basis has to be computed. Thus, we remove the inactive constrains only when the number of constraints in the current LP problem exceeds $T$, for some integer $T$. Note that this differs from the algorithm called MALP-B in \cite{zha12}, where the inactive constraints are removed in each iteration.

Another way to decrease the running-time of the ZS algorithm in some cases is to terminate the ZS decoder prematurely as soon as the objective value exceeds the current upper bound $\tau$ in Algorithm~\ref{alg:sse}. In that event the objective function value cannot possibly be improved below the current node, and it can be skipped immediately.

\subsection{Tradeoff Between Tightness and Speed of the ZS Algorithm}\label{sec:LPcomplexity}
The cut-search procedure used in the ZS decoding algorithm yields tight lower bounds on the MLD solution at the cost of a high number of cuts and thus increased processing time spent in the LP solver and for Gaussian elimination (see Section~\ref{sec:ZS}). In two different ways, a tradeoff between speed and tightness can be realized. First, limit the maximum number of times $R$ that the search for redundant parity-check cuts is applied, and secondly, only add a cut if the \emph{cutoff}, i.\,e., the distance between the current (infeasible) solution and the cutting hyperplane, exceeds a fixed quantity $\gamma>0$.

Our numerical experiments have shown that both approaches help to significantly reduce the running-time complexity of our algorithm. Additionally, for the first approach, it has proven helpful to use a higher value $R^{\text{bb}}$ in those iterations where a best-bound node has been selected (cf. Section~\ref{sec:nodeselection}).

\subsection{Special Case: MLD Performance Simulation}\label{sec:curve}
For benchmarking purposes we are only interested in the actual MLD curve, in which case the MLD algorithm can be simplified. First, since the underlying code is always linear, the error probability of MLD is independent of the actual transmitted codeword, thus we can always, without loss of generality, transmit the all-zero codeword. Furthermore, when the all-zero codeword is transmitted and a codeword $\mathbf{c}$ with objective value $\psi_{\boldsymbol{\lambda}}(\mathbf{c}) < 0 = \psi_{\boldsymbol{\lambda}}(\boldsymbol{0})$ has been identified, the search can be terminated, since any ML decoder would also fail on this received LLR vector $\boldsymbol{\lambda}$.

\begin{table*}[t]
\scriptsize \centering \caption{Numerical comparison of our proposed algorithm and CPLEX for several codes  for different values of the SNR on the AWGN channel. $T_{\rm avg}$ is the average decoding time per frame in seconds, and $N_{\rm avg}$ is the average number of nodes (per frame) processed by the branch-and-bound algorithms. The numbers in the parentheses are for CPLEX. In all cases but the first we used all-zero decoding as described in Section~\ref{sec:curve}.} \label{tab:comp}
\def\Hline{\noalign{\hrule height 2\arrayrulewidth}}
\vskip -3.0ex 
\begin{tabular}{lccccccc}
\Hline \\ [-2.0ex]
 & SNR in \si{\decibel} & $1.0$ & $1.5$ & $2.0$ & $2.5$ & $3.0$ & $3.5$ \\
\hline
\\ [-2.0ex] \hline  \\ [-2.0ex]
\multirow{2}{*}{$(155,64)$ Tanner code \cite{tan01}}  & $T_{\rm avg}$ &
  $0.81$ ($9.49$) &
  $0.24$ ($2.95$) &
  $0.05$ ($0.63$) &
  $0.014$ ($0.17$) &
  $0.005$ ($0.095$) &
  $0.004$ ($0.086$)\\ 
& $N_{\rm avg}$ &
  $51$ ($4795$) &
  $15$ ($1816$) &
  $3.5$ ($370$) &
  $1.4$ ($61$) &
  $1.04$ ($5.4$) &
  $1.004$ ($0.3$) \\
\multirow{2}{*}{$(155,64)$ Tanner code \cite{tan01} (all-zero)}  & $T_{\rm avg}$ &
  $0.24$ ($3.23$) &
  $0.11$ ($1.16$) &
  $0.025$ ($0.28$) &
  $0.006$ ($0.06$) &
  $0.001$ ($0.02$) &
  $0.0005$ ($0.01$)\\ 
& $N_{\rm avg}$ &
  $14$ ($2799$) &
  $6.6$ ($963$) &
  $2.1$ ($210$) &
  $1.18$ ($38$) &
  $1.05$ ($4.5$) &
  $1.002$ ($0.3$) \\
\multirow{2}{*}{$(204,102)$ MacKay code \cite{mac-web} (all-zero)} & $T_{\rm avg}$ &
  $2.2$ ($14.6$) &
  $0.73$ ($4.7$) &
  $0.15$ ($0.83$) &
  $0.02$ ($0.12$) &
  $0.003$ ($0.03$) &
  $0.0005$ ($0.018$)\\ 
& $N_{\rm avg}$ &
  $90$ ($12364$) &
  $30.5$ ($3421$) &
  $6.5$ ($573$) &
  $1.66$ ($61$) &
  $1.04$ ($4.9$) &
  $1.003$ ($0.19$) \\
\multirow{2}{*}{$(127,85)$ BCH code (all-zero) }  & $T_{\rm avg}$ &
  $86$ (--) &
  $67$ (--) &
  $33$ (--) &
  $9$ (--) &
  $2.2$ (--) &
  $0.29$ ($3.5$)\\ 
& $N_{\rm avg}$ &
  $7617$ (--) &
  $5855$ (--) &
  $2549$ (--) &
  $655$ (--) &
  $159$ (--) &
  $19$ ($4132$) 
\end{tabular}
\vspace{-2ex}
\end{table*}

\section{Minimum Distance Computation}
The MLD problem is closely related to the computation of the minimum distance $d_{\min}$ of a code as follows. If the all-zero codeword is explicitly forbidden, then an MLD algorithm with the input $\boldsymbol{\lambda} = \boldsymbol{1}$ will output a codeword of minimum weight:
\begin{equation} \label{eq:MD}
d_{\min}(\mathcal C)=\min_{\mathbf{c} \in \mathcal{C} \setminus \{\boldsymbol{0}\}} \psi_{\boldsymbol{1}}(\mathbf{c}) = \min_{\mathbf{c} \in \conv(\mathcal{C}\setminus\{\boldsymbol 0\})} \psi_{\boldsymbol{1}}(\mathbf{c}).
\end{equation}
Our proposed decoding algorithm can be modified to exclude the all-zero codeword by the following changes:
\begin{enumerate}
  \item Extend the condition in Step~\ref{line:integrality} of Algorithm~\ref{alg:sse} to \enquote{$\hat{\mathbf{p}}$ is integral and $\hat{\mathbf{p}} \neq \mathbf{0}$}, which avoids decoding to the all-zero codeword.
  \item In the order-$i$ re-encoding performed in Algorithm~\ref{alg:ZS}, exclude $\mathbf{0}$ from the set of candidate codewords.
\end{enumerate}
Moreover, note that all feasible solutions (i.\,e., codewords) of \eqref{eq:MD} have an integral objective value. This allows us to change the right hand side in Steps \ref{line:mainloop}, \ref{line:prune1}, and \ref{line:prune2} to $\tau-1+\varepsilon$, for a small $\varepsilon>0$, since $\left\lceil\bar\psi_{\min,\mathbf 1}^{(F)}\right\rceil = \tau$ implies that $\psi_{\min,\mathbf 1}^{(F)} \geq \tau$.

\section{Numerical Results}\label{sec:results}
\begin{figure}
\centering
\pgfplotsset{grid style={dotted}}
\begin{tikzpicture}
  \begin{semilogyaxis}[
          xlabel=$E_b/N_0 (\si{\decibel})$,
          ylabel=frame error rate,
          grid=both,
          height=6cm,
          width=0.95\columnwidth,
          legend style={legend pos=south west,font=\scriptsize}]
          
      \addplot[mark=diamond*,color=red] plot coordinates {
          (1, 4.05e-1)
          (1.5, 1.77e-1)
          (2.0, 5.74e-2)
          (2.5, 1.5e-2)
          (3.0, 2.14e-3)
          (3.5, 1.98e-4)
      };
      \addlegendentry{$(127,85)$ BCH code}
      
      \addplot[mark=triangle*,color=black] plot coordinates {
          (1, 1.19e-1)
          (1.5, 4e-2)
          (2, 6.57e-3)
          (2.5, 6.02e-4)
          (3, 7.68e-5)
          (3.5, 1.56e-5)
      };
      \addlegendentry{$(204,102)$ MacKay code}
      
      \addplot[mark=x,color=green!50!black] plot coordinates {
          (1, 6.89e-2)
          (1.5, 1.79e-2)
          (2.0, 4.75e-3)
          (2.5, 7.75e-4)
          (3.0, 6.37e-5)
          (3.5, 4.11e-6)
      };
      \addlegendentry{$(155,64)$ Tanner code}
\end{semilogyaxis}
\end{tikzpicture}\vspace{-2ex}
\caption{MLD performance of the codes considered in this paper.}\label{fig:mlcurves}\vspace{-2ex}
\end{figure}
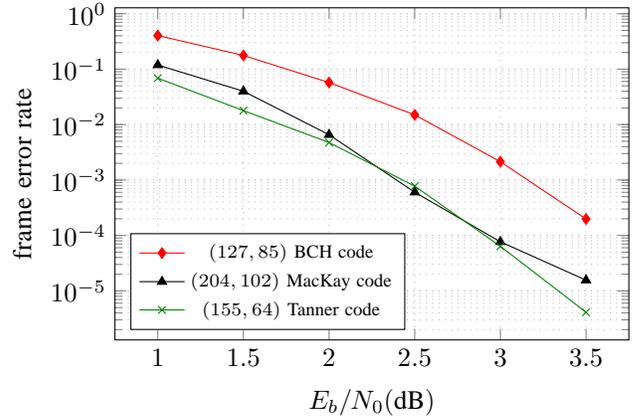
In this section, we present some numerical results for our proposed MLD algorithm, with all the improvements outlined above in Section~\ref{sec:improvements}, for several codes on the AWGN channel. We have used order-$2$ re-encoding ($i=2$ in Algorithm~\ref{alg:ZS}), and the open-source GLPK library \cite{glpk} to solve the LP problems. The following set of parameters was heuristically found to perform well for all codes: $M=30$, $\delta=2$, 
$T=100$, 
$R=5$, $R^{\text{bb}}=100$, and $\gamma=0.2$. 

As a benchmark, we use the CPLEX IP solver \cite{CPLEX126}, with the IP formulation named IPD1 in \cite{Tanatmis+10NumericalComparison} which was found to be most efficient in that paper. In case of all-zero decoding, we configured CPLEX to terminate as soon as a codeword with objective value below zero was found, mimicking the adaptions of our algorithm described in Section~\ref{sec:curve}.

We compare the algorithms with respect to both (single-core) average CPU time $T_{\text{avg}}$ and average number $N_{\text{avg}}$ of branch-and-bound nodes processed per frame. For our algorithm, $N_{\text{avg}}$ equals the number of times the main loop (Step~2 of Algorithm~\ref{alg:sse}) is processed, while for CPLEX we report the attribute \enquote{number of processed nodes}. Note that the latter drops below one for high SNR, which is probably due to CPLEX' presolve strategy that establishes optimality in some cases without ever starting the branch-and-bound procedure.

All calculations were performed on a desktop PC with an Intel Core i5-3470 CPU (\SI{3.2}{\giga\hertz}) and \SI{8}{\giga\byte} of RAM. 

\subsection{Maximum-Likelihood Decoding}
A comparison of CPLEX and our MLD algorithm, for the different codes outlined below, is given in  Table~\ref{tab:comp}, both in terms of $T_{\rm avg}$ and $N_{\rm avg}$. The numbers in the parentheses are for CPLEX; a dash indicates that CPLEX was not able to decode a sufficient number of frames without running out of memory. The corresponding MLD performance curves are plotted in Fig.~\ref{fig:mlcurves}. For the curves, we have counted 100 erroneous frames for each simulation point.

The $(155,64)$ Tanner code from \cite{tan01}
is often used as a benchmark code, and was also considered in \cite{zha12}. For all SNRs, the ZS decoding algorithm showed a performance loss compared to the MLD curve \cite{zha12}. 
As can be seen from Table~\ref{tab:comp}, our algorithm is more than 11 times as fast as CPLEX for an SNR of \SI{1.0}{\decibel}. For higher values of the SNR  our proposed algorithm is even faster compared to CPLEX. In the case of all-zero decoding, both algorithms are faster by a factor of $2$ to $3$, while the relative performance gain by our algorithm remains roughly the same.


The second example is a $(3,6)$-regular $(204,102)$ LDPC code taken from the online database of sparse graph codes from MacKay's website \cite{mac-web} (called 204{.}33{.}484 there). As can be seen from Table~\ref{tab:comp}, also for this code, our algorithm is significantly faster than CPLEX for all  simulated SNRs. 

In order to evaluate the performance of our algorithm for dense codes, Table~\ref{tab:comp} includes results for the $(127,85)$ BCH code. CPLEX was not able to decode a significant number of frames for this code, and to our knowledge the MLD curve, as presented in Fig.~\ref{fig:mlcurves}, was previously unknown.
%
%

\begin{table}
\scriptsize \centering \caption{Numerical results for minimum distance computation.} \label{tab:dmin}
\def\Hline{\noalign{\hrule height 2\arrayrulewidth}}
\vskip -3.0ex 
\begin{tabular}{lccccccc}
\Hline \\ [-2.0ex]
 & $d_{\min}$ & $T^{\text{MLD}}$  & $T^{\text{CPLEX}}$ & $N^{\text{MLD}}$ & $N^{\text{CPLEX}}$ \\
\hline
\\ [-2.0ex] \hline  \\ [-2.0ex]
$(155, 64)$ Tanner code & \si{20} & \SI{137}{\second} & \SI{3682}{\second} & \si{42785} & \si{21842224} \\
 $(204,102)$ MacKay code & \si{8} & \SI{1.6}{\second} & \SI{11.49}{\second} & \si{371} & \si{44830} \\
 $(408,204)$ MacKay code & \si{14} & \SI{152}{\second} & \SI{6893}{\second} & \si{9345} & \si{936570}
\end{tabular}
\vspace{-3ex}
\end{table}

\subsection{Minimum Distance Computation}
In the case of computing the minimum distance, we used different values for some of the parameters, namely $M=120$, $R=1$, $R^{\text{bb}}=1$, and $\gamma=0.3$. Results are shown in Table~\ref{tab:dmin}, which additionally contains the $(408,204)$ MacKay code (named 408{.}33{.}844 at the website \cite{mac-web}) that was used also in \cite{zha12}. Note that in case of the $(155,64)$ Tanner code, we can exploit the symmetry and fix $c_0=1$ before starting the algorithm. We compare our algorithm to CPLEX with the same formulation as for MLD and the additional constraint $\sum_{i=0}^{n-1}c_i \geq 1$ to exclude the all-zero codeword.\footnote{As a remark, for the $(408, 204)$ MacKay code we have used the previous CPLEX 12.5 instead of 12.6; apparently there is a bug in the latter, causing it to output a $d_{\min}$ of $20$ instead of the correct value $14$.}



\bibliographystyle{IEEEtran}

\begin{thebibliography}{10}
\providecommand{\url}[1]{#1}
\csname url@samestyle\endcsname
\providecommand{\newblock}{\relax}
\providecommand{\bibinfo}[2]{#2}
\providecommand{\BIBentrySTDinterwordspacing}{\spaceskip=0pt\relax}
\providecommand{\BIBentryALTinterwordstretchfactor}{4}
\providecommand{\BIBentryALTinterwordspacing}{\spaceskip=\fontdimen2\font plus
\BIBentryALTinterwordstretchfactor\fontdimen3\font minus
  \fontdimen4\font\relax}
\providecommand{\BIBforeignlanguage}[2]{{%
\expandafter\ifx\csname l@#1\endcsname\relax
\typeout{** WARNING: IEEEtran.bst: No hyphenation pattern has been}%
\typeout{** loaded for the language `#1'. Using the pattern for}%
\typeout{** the default language instead.}%
\else
\language=\csname l@#1\endcsname
\fi
#2}}
\providecommand{\BIBdecl}{\relax}
\BIBdecl

\bibitem{ber78}
E.~R. Berlekamp, R.~J. McEliece, and H.~C.~A. {van Tilborg}, ``On the inherent
  intractability of certain coding problems,'' \emph{IEEE Trans.~Inf.~Theory},
  vol.~24, no.~3, pp. 384--386, May 1978.

\bibitem{Tanatmis+10NumericalComparison}
A.~Tanatmis, S.~Ruzika, M.~Punekar, and F.~Kienle, ``Numerical comparison of
  {IP} formulations as {ML} decoders,'' in \emph{Proc. {IEEE} Int. Conf.
  Commun. (ICC)}, Cape Town, South Africa, May 2010.

\bibitem{CPLEX126}
``{IBM} {ILOG} {CPLEX} {O}ptimization {S}tudio,'' Commercial Software Package,
  2013, version 12.6.

\bibitem{ros12}
E.~Rosnes, {\O}.~Ytrehus, M.~A. Ambroze, and M.~Tomlinson, ``Addendum to
  \enquote*{{A}n efficient algorithm to find all small-size stopping sets of
  low-density parity-check matrices},'' \emph{IEEE Trans.~Inf.~Theory},
  vol.~58, no.~1, pp. 164--171, Jan. 2012.

\bibitem{ros09}
E.~Rosnes and {\O}.~Ytrehus, ``An efficient algorithm to find all small-size
  stopping sets of low-density parity-check matrices,'' \emph{IEEE
  Trans.~Inf.~Theory}, vol.~55, no.~9, pp. 4167--4178, Sep. 2009.

\bibitem{fel05}
J.~Feldman, M.~J. Wainwright, and D.~R. Karger, ``Using linear programming to
  decode binary linear codes,'' \emph{IEEE Trans.~Inf.~Theory}, vol.~51, no.~3,
  pp. 954--972, Mar. 2005.

\bibitem{tag07}
M.~H. Taghavi and P.~H. Siegel, ``Adaptive methods for linear programming
  decoding,'' \emph{IEEE Trans.~Inf.~Theory}, vol.~54, no.~12, pp. 5396--5410,
  Dec. 2008.

\bibitem{zha12}
X.~Zhang and P.~H. Siegel, ``Adaptive cut generation algorithm for improved
  linear programming decoding of binary linear codes,'' \emph{IEEE
  Trans.~Inf.~Theory}, vol.~58, no.~10, pp. 6581--6594, Oct. 2012.

\bibitem{tan01}
R.~M. Tanner, D.~Sridhara, and T.~Fuja, ``A class of group-structured {LDPC}
  codes,'' in \emph{Proc. Int. Symp. Commun. Theory and Appl. (ISCTA)},
  Ambleside, England, Jul. 2001.

\bibitem{ksc01}
F.~R. Kschischang, B.~J. Frey, and H.-A. Loeliger, ``Factor graphs and the
  sum-product algorithm,'' \emph{IEEE Trans.~Inf.~Theory}, vol.~47, no.~2, pp.
  498--519, Feb. 2001.

\bibitem{fos95}
M.~P.~C. Fossorier and S.~Lin, ``Soft-decision decoding of linear block codes
  based on ordered statistics,'' \emph{IEEE Trans.~Inf.~Theory}, vol.~41,
  no.~5, pp. 1379--1396, Sep. 1995.

\bibitem{hel12}
M.~Helmling, S.~Ruzika, and A.~Tanatmis, ``Mathematical programming decoding of
  binary linear codes: Theory and algorithms,'' \emph{IEEE Trans.~Inf.~Theory},
  vol.~58, no.~7, pp. 4753--4769, Jul. 2012.

\bibitem{koe05}
\BIBentryALTinterwordspacing
P.~O. Vontobel and R.~Koetter, ``Graph-cover decoding and finite-length
  analysis of message-passing iterative decoding of {LDPC} codes,''
  arXiv:cs/0512078 [cs.IT], Dec. 2005. [Online]. Available:
  \url{http://arxiv.org/abs/cs.IT/0512078/}
\BIBentrySTDinterwordspacing

\bibitem{faigle10}
U.~Faigle, W.~Kern, and G.~Still, \emph{Algorithmic Principles of Mathematical
  Programming}.\hskip 1em plus 0.5em minus 0.4em\relax Kluwer Academic
  Publishers, 2010, vol.~24.

\bibitem{mac-web}
\BIBentryALTinterwordspacing
D.~J.~C. MacKay, encyclopedia of sparse graph codes. [Online]. Available:
  \url{http://www.inference.phy.cam.ac.uk/mackay/codes/data.html}
\BIBentrySTDinterwordspacing

\bibitem{glpk}
\BIBentryALTinterwordspacing
``{GNU} {L}inear {P}rogramming {K}it ({GLPK}),'' Software Library, version
  4.52. [Online]. Available: \url{http://www.gnu.org/software/glpk}
\BIBentrySTDinterwordspacing

\end{thebibliography}
{\small \itemsep 10ex

}

\end{document}